\documentclass[11pt,a4paper,reqno]{amsart}
 \usepackage{amsfonts,amsthm, amscd, epsfig, amsmath, amssymb,enumerate}
 \usepackage[english]{babel}
\numberwithin{equation}{section}

\title{Mott law as upper bound for a random walk in a random environment.}

\author{A. Faggionato}
\address{Alessandra Faggionato. Dipartimento di Matematica ``G. Castelnuovo", Universit\`a   ``La
  Sapienza''. P.le Aldo Moro  2, 00185  Roma, Italy. e--mail:
  faggiona@mat.uniroma1.it}

\author{P. Mathieu}
\address{Pierre Mathieu.  Centre de Math\'ematiques et d'Informatique (CMI), Universit\'e   de Provence.
39 rue Joliot Curie, 13453 Marseille cedex 13, France.
 e--mail:
pierre.mathieu@cmi.univ-mrs.fr
}

\newtheorem{theo}{Theorem}

\newtheorem{prop}{Proposition}
\newtheorem{lemma}{Lemma}

\newtheorem{rem}{Remark}

\newcommand{\XX}{{\mathbb X}}

\newcommand{\NN}{{\mathbb N}}
\newcommand{\RR}{{\mathbb R}}

\newcommand{\ZZ}{{\mathbb Z}}

\newcommand{\Pp}{{\mathcal P}}
\newcommand{\Qq}{{\mathcal Q}}
\newcommand{\PP}{{\bf P}}

\newcommand{\EE}{{\bf E}}
\newcommand{\Bb}{{\mathcal B}}
\newcommand{\Gg}{{\mathcal G}}

\newcommand{\Ee}{{\mathcal E}}

\newcommand{\Tr}{\mbox{\rm Tr}}

\newcommand{\Nn}{{\mathcal N}}
\newcommand{\Cc}{{\mathcal C}}
\newcommand{\Vv}{{\mathcal V}}

\newcommand{\II}{{\mathbb I}}
\let\a=\alpha \let\b=\beta   \let\d=\delta  \let\e=\varepsilon
 \let\g=\gamma     \let\k=\kappa

  \let\z=\zeta
\let\D=\Delta   \let\G=\Gamma

\begin{document}

\maketitle
\
\begin{abstract}
We consider a  random walk  on the support of an ergodic   simple
point process on $\RR^d$, $d\geq 2$, furnished with independent
energy marks. The  jump rates of the random walk decay exponentially
in the jump length and depend on the  energy marks via a
Boltzmann--type factor. This is an effective model for the
phonon--induced hopping of electrons in disordered solids in the
regime of strong Anderson localization. Under some technical
assumption on the point process we prove an upper bound for the
diffusion matrix of the random walk in agreement with Mott law. A
lower bound for $d\geq 2$  in agreement with Mott law was proved in
\cite{FSS}.

\smallskip




 \end{abstract}

\section{Introduction}

\subsection{Physical motivations}
 Phonon-assisted electron transport in  disordered solids in which the
Fermi level (set equal to $0$ below) lies in a region of  strong
Anderson localization can be modeled
 by    Mott variable--range hopping of  the following   interacting particle
system in a  random environment \cite{SE}. The environment is given
by $\xi:= \bigl( \{x_i\}, \{ E_{i}\} \bigr)$, where $\{x_i\}$ is an
infinite
  and locally finite set of points in   $\RR^d $ such that  each point
$x_i$ is labeled by an energy mark $E_i$ belonging to some finite
interval.
Given $\xi$, particles  can lie only at points of $\{x_i\}$ and
  perform   random walks on $\{x_i\}$ with hard--core interaction.
The probability rate for a jump of a particle at  $x$ to the vacant
site $y$, with $x\not =y$ in $\{x_i\}$, is given by
\begin{equation}\label{mariano}
  r_{x,y}(\xi)= r_0 \exp\left\{ -2 |x-y|/\ell_*   -\b
\{E_y-E_x\}_+\right\}\, ,
\end{equation}
where   $E_z$ is defined by $E_i$ for $z=x_i$ and
  $\{E_y-E_x\}_+= \max\{ E_y-E_x, 0\}. $  Above $|\cdot |$ denotes the
  Euclidean norm in $\RR^d$,  $\b =1/(kT)$, $\ell_*$
denotes the localization radius of wavefunctions and $r_0$ is a constant depending on
the material but which depends only weakly on $\b$, $|x-y|$ and the
energies $E_x,E_y$. Without loss of generality, in what follows we
set $\ell_*=2$, $r_0=1$.

The disorder of the solid is   modeled by the  randomness of the
environment. The points $\{x_i\}$  correspond to  the {\em
impurities} of the disordered solid and the electron Hamiltonian has
exponentially localized quantum eigenstates with localization
centers $x_i$ if the corresponding energies $E_i$ are close to the
Fermi level. The random set $\{x_i\}$ is a stationary simple point
process with finite intensity (i.e.\ with finite mean number of
points in finite boxes) and its stationarity reflects the
homogeneity of the medium. Conditioned to $\{x_i\}$, the energy
marks $\{E_i\}$ are supposed to be i.i.d. random variables taking
value in a finite interval (set in what follows equal to $[-1,1]$)
with common distribution $\nu$, such that $\nu(dE)\sim c\,|E|^\a dE
$, $|E|\ll 1$, for a suitable nonnegative exponent $\a$. The
independence of the energy marks is compatible with Poisson level
statistics, which is a general rough indicator for the localization
regime and has been proven to hold for an Anderson model \cite{Min}.
The exponent $\alpha$ allows to model a possible Coulomb pseudogap
in the density of states \cite{SE}. In particular, the physically
relevant possible values of $\a$ are $0$ and $d-1$, the latter
corresponding to the Coulomb pseudogap.

The DC  conductivity matrix $\sigma (\b)$, measuring
  the linear response of the solid to a uniform external
electric field, would vanish if it were not for the lattice
vibrations (phonons) at nonzero temperature.  For an isotropic solid
at low temperature and with dimension $d\geq 2$,   Mott law
\cite{MOTT1}, \cite{MOTT2} predicts  that the conductivity decays
exponentially as
\begin{equation}\label{mottlaworigin}
\sigma (\b ) \sim  \exp\left\{-c \,  \b
^{\frac{\a+1}{\a+1+d}}\right\} \, \II\,, \qquad \b \gg 1\,,
\end{equation}
where $c$ is a  $\b$--independent positive constant  and  $\II$
denotes the identity matrix. The original derivation of
(\ref{mottlaworigin}) is based on an heuristic optimization
argument,
 alternative and more robust derivations  have been proposed
  after that (see \cite{AHL}, \cite{SE} and references
therein).
 Finally, we point out that due to the Einstein relation between
 $\sigma(\b)$ and
the bulk diffusion matrix $D_{\text{bulk}}(\b)$  \cite{Sp},
 (\ref{mottlaworigin}) is equivalent to the asymptotic behavior
\begin{equation}\label{mottlaworiginbis}
D_{\text{bulk}} (\b ) \sim  \exp \left\{-c \,   \b
^{\frac{\a+1}{\a+1+d}}\right\} \, \II\,, \qquad \b\gg 1 \,.
\end{equation}

\medskip

 A mean field version of Mott variable--range hopping
  at small temperature
  is given by the continuous--time random walk $X_t^\xi$  on $\{x_i\}$ such that
  the
   jump from $x$ to $y$, $x\not = y$ in $\{x_i\}$, has probability
   rate
   given by
\begin{equation}
\label{eq-rates} c_{x,y}(\xi)= \exp\left\{-
|x-y|-\beta(|E_x-E_y|+|E_x|+|E_y|)/2 \right\}\,.
\end{equation}
 
 We will call
$X_t^\xi$  {\em Mott variable--range random walk}.

 The choice  of the transition rates $c_{x,y}(\xi)$ comes from the
 fact that,  at small temperature,
 \begin{equation}\label{equivalenza_c_r}
c_{x,y}(\xi)=  r_{x,y} (\xi) \, \mu (\eta_x=1 )\, \mu (\eta _y =0)\,
(1+o(1)) \,, \qquad \b\gg 1\,,
  \end{equation}
 where $\eta_x$ is the particle number at site
$x$ in the above particle  system, while $\mu$ is the Gibbs measure
of the particle system w.r.t. the Hamiltonian $H=\sum _i
E_i\eta_{x_i}$  with  zero Fermi level, i.e. $\mu$ is the
product measure on $\{0,1\}^{\{x_i\}}$ such that
 $$
 \mu(\eta_{x_i}=1) = e^{-\b
E_i}/ (1+ e^{-\b E_i})\, . $$ In (\ref{equivalenza_c_r}), the error
term is negligible as $\b\rightarrow \infty$ uniformly  in $x,y$
belonging to a finite volume (as in realistic solids), for a typical
environment $\xi$.
 Another
derivation of the mean field version (\ref{eq-rates}) from the
original Mott variable range hopping (\ref{mariano}) has been
obtained  in \cite{MA} (see also \cite{AHL}[Section IV])  by
reduction to a random resistor network .

The analogous  of Mott law (\ref{mottlaworiginbis}) for Mott
variable--range random walk is given by
\begin{equation}\label{mottmeanfield1}
D (\b ) \sim  \exp \left\{ -c \,   \b ^{\frac{\a+1}{\a+1+d}}\right\}
\II\,, \qquad \b \gg 1\,,
\end{equation}
where  $D(\b)$ denotes the diffusion matrix of the random walk. For
dimension $d\geq 2$, a lower bound of $D(\b)$ in agreement with
(\ref{mottmeanfield1}) has been recently proven in \cite{FSS}. The
present work addresses to the problem of
rigorously  deriving      an upper bound of $D(\b)$ in agreement with
(\ref{mottmeanfield1}). The one dimensional case present special
features and rigorous results on the corresponding Mott law  have
been obtained in \cite{CF2}.

\subsection{Model and results}
Let us give a precise definition of  Mott variable--range random walk $X_t^\xi$, generalizing 
the choice of jump rates (\ref{eq-rates}).   The  environment
$\xi=\bigl(\{x_i\}, \{E_i\}\bigr)$   is defined
 as follows. Let $\{x_i\}$ be a simple point process, i.e.\ a random
locally finite subset of $\RR^d$, 
 and its law is the Palm distribution $\hat\Pp_0$ associated to the law $\hat
\Pp$ of a stationary simple point process on $\RR ^d$ with finite
intensity. Given $\{x_i\}$, the energy marks $\{E_i\}$ are i.i.d.
random variables having value in  $[-1,1]$ and common law $\nu$ (restrictions
 on $\nu$
will be specified later).  The law  $\Pp_0$ of the environment  is
the so called $\nu$--randomization of the Palm distribution $\hat
\Pp_0$ associated to $\hat \Pp$. 

 The reason for considering  Palm distributions is the 
following: in order to make the random walk start
 at a fixed point, taken equal to the
origin below,  we condition to contain the 
origin  the
stationary point process  of impurities introduced in the previous subsection.  
 As discussed in Section \ref{basicfacts}, if the stationary process has finite intensity
and law  $\hat \Pp$, the law of the resulting process 
 is given by the Palm distribution $\hat \Pp _0$
associated to $\hat \Pp$.

In what follows, we  write $\xi$ for a generic
locally finite subset of $\RR^d\times[-1,1]$
such that  $\xi$ has at most one point in each  fiber 
$ \{x\}\times [-1,1]$,
 and we write  $\hat \xi$ for a
generic locally finite subset of $\RR^d$. Both $\xi$ and $\hat \xi$
can be identified with the counting measures $\sum _{(x,E)\in \xi } \d_{(x,E)}$ and $\sum
_{x\in \hat \xi}\d_x$  respectively. Moreover, when there is no
ambiguity,  given   $\xi$ its spatial projection on $\RR^d$ will be denoted by $\hat \xi$.
We finally  recall that  the $\k$--moment $\rho_\k$ of $\hat\Pp$, $\k>0$, is defined as  
$$\rho_\k:=\EE_{\hat \Pp }\left( |\hat \xi \cap [0,1]^d|^\k\right)\,.$$
 Then $\rho:=\rho_1$ is  called the
intensity of the process $\hat \Pp$.

\smallskip

Given a realization of the environment $\xi$, Mott variable--range
random walk $X_t^\xi$  is defined as the continuous--time random walk  
on  $\hat\xi= \{x_i\}$ starting at the origin and  jumping  from $x$ to $y$, $x\not =y$ in
$\{x_i\}$, with probability rate
\begin{equation}\label{eq_finale}
c_{x,y}(\xi)=  \exp\left\{-|x-y|-\b u(E_x,E_y)\right\}\,,
  \end{equation}
where the function $u$ satisfies
\begin{equation}\label{uuu}
\k_1 (|E_x|+|E_y|) \leq u(E_x,E_y)\leq \k_2( |E_x|+|E_y|)
\end{equation}
for some positive constants $\k_1 \leq \k_2$.  To simplify the
notation, it is convenient to set $c_{x,x}(\xi)\equiv 0$ for all
$x\in \{x_i\} $.

The law $\mathbb{P} ^\xi$ of 
$X^\xi_t$ is characterized   by the following identities:
\begin{align*}
& \mathbb{P}^\xi (X^\xi _0=0)=1\,,\\
 & \mathbb{P}^\xi  ( X^\xi_{t+dt}
=y\,|\, X^\xi _t=x) = c_{x,y}(\xi) dt +
o(dt),\qquad t\geq 0,\, \;x\neq y\,,\\
&   \mathbb{P}^\xi  ( X^\xi _{t+dt} =x\,|\, X^\xi _t=x) =
1-\sum _z c_{x,z}(\xi) dt + o(dt),\qquad t\geq 0\,.
\end{align*}

Equivalently, the dynamics of  $X^\xi_t $ can be described as
follows: after arriving at site $x$ the particle waits an
exponential time with
  parameter
  \begin{equation}
   \lambda _x(\xi)=\sum _z c_{x,z }(\xi)\,,
  \label{lao}\end{equation}
and then jumps to site $y$, $y\not= x $, with probability
  \begin{equation}\label{eq_prob}
  \frac{c_{x,y}(\xi)}{\lambda _{x} (\xi) }\,.
  \end{equation}
 By standard methods (see e.g.\ \cite{Br}, \cite{FSS}[Appendix A]),
    one can check that the random walk
     $X^\xi_t$ is well--defined for
  $\Pp_0$--a.a.\ $\xi$  as soon as $\hat \Pp$ is ergodic w.r.t.
  spatial translations and
$\EE_{\Pp_0}\left( \lambda _0 (\xi) \right)<\infty$. As proven in
\cite{FSS} (see also Lemma \ref{techn_lemma} below) this last
condition   is equivalent to require that $\rho_2 <\infty$.

\smallskip

Assuming $\hat \Pp$ to be ergodic, $\rho_{12}$ to be finite
  and
under  some  additional technical assumption  on the law of the environment
  $\Pp_0$, in
\cite{FSS} the authors prove that the diffusively rescaled  random
walk $X^\xi$ converges in  $\Pp_0$--probability to a Brownian motion
whose covariance matrix coincides with $D(\b)$, where $D(\b)$ is
the diffusion matrix
of $X^\xi$ defined as the unique symmetric matrix such that
\begin{equation}
\label{eq-Diffmatrix} (a\cdot D(\b) a) =
\lim_{t\to\infty}\,\frac{1}{t} \,\EE_{\Pp_0}\left(\EE_{\PP^\xi_0 }
\left((X_t^\xi \cdot a)^2 \right)\right) \,, \qquad a\in\RR^d\,.
\end{equation}
Moreover, they prove
 the following variational characterization of  $D(\b)$:
 \begin{equation}
\label{sereno} \left(a, D(\b) \, a\right) = \inf_{f\in
L^\infty(\Pp_0)}  \int \Pp_0(\xi) \int \hat{\xi} (dx)\,
c_{0,x}(\xi)\, \left( a\cdot x-\nabla_x f(\xi) \right)^2 , \qquad
a\in\RR^d\, ,
\end{equation}
where
 the set $\hat \xi$ is  defined as the spatial projection of $\xi$, i.e.\  $\hat \xi=\{x_i\}$,
  and is identified with the counting measure $\sum _i \d_{x_i}  $.  Moreover,
 the gradient $\nabla _x f(\xi)$ is defined as
$$
\nabla_x f(\xi)=f(S_x\xi) - f(\xi) ,\qquad S_x \xi:=\left\{ (x_i-x,
E_i )\right\}.
$$
 In addition,  for  $d\geq 2$  and assuming that
\begin{equation}\label{lowerbound}
  \nu ( [-E,E ])\geq c_0 |E|^{\a+1} \,,  \qquad \forall E\in [-1,1]\, , 
\end{equation}
   for some positive
constant $c_0$ and some exponent $\a\geq 0$,
  the authors  prove  a lower bound on $D(\b)$:  for $\b$ large enough it holds
 \begin{equation}\label{Dlowerbound}
 D(\b)\geq  c_1 \b ^{-c_2 }  \exp \left\{ -C\, \b ^{\frac{\a+1}{\a+1+d}}\right\} \II
 \, ,
\end{equation}
 where $c_1, c_2, C $ are  positive constants independent of $\b$. The above
bound is in agreement with Mott law (\ref{mottmeanfield1}).

We  note that the requirement $\a>0$ in
 \cite{FSS}[Theorem 1]  is due to a typing error. Moreover  
 it is simple to check that, although
  in \cite{FSS} the transition rates $c_{x,y}(\xi)$ are defined as
  in (\ref{eq-rates}), all the results in \cite{FSS} remain true for transition rates $c_{x,y}(\xi)$
   defined as
  in (\ref{eq_finale}) and  assuming only that  $\a>-1$.

\medskip

Our main result consists in an   upper bound for $D(\b)$ in
agreement with Mott law. Roughly, we claim that
 if  (\ref{lowerbound}) holds with inverted sign, then  also (\ref{Dlowerbound})
  remains valid with inverted sign. In order to precisely state  our
  technical assumptions we fix some notation.

Given  $p\in [0,1]$ and    a simple point process
  with law $\hat \Pp$,
its \emph{$p$--thinning} is the simple point process obtained as
follows: for each   realization $\hat \xi$ of the process with law
$\hat \Pp$ erase each point independently with probability $1-p$. We
will write $\hat \Pp ^{(p)}$ for the law of the $p$--thinning of
$\hat \Pp$.

Finally,
    we recall that  the {\sl Poisson point process}   on $\RR^d$
with intensity $\rho >0$ is a random locally finite subset  $\hat
\xi \subset \RR^d$ such that (i) for any $A\subset \RR^d$ Borel and
bounded,   the cardinality $ \hat \xi (A )$ is a Poisson random
variable with expectation $ \rho \,\ell (A)$ where $\ell (A)$ is the
Lebesgue measure of $A$; (ii)  for any disjoint Borel subsets
$A_1,\dots, A_n\subset \RR^d $, $ \hat \xi (A_1),\dots,
  \hat \xi ( A_n)$ are independent random variables. We denote
by $\hat \Pp _\rho$ the law of the Poisson point process with
intensity $\rho$. The process $\hat \Pp _\rho$ is stationary. 

\bigskip

We can finally state our main result:

\begin{theo}\label{teo-Mottup}
Let $\Pp_0$ be the  $\nu$--randomization of the  Palm distribution  $\hat \Pp_0$   associated to 
 a  stationary simple  point process   on
$\RR^d$, $d\geq 2$, with law $\hat \Pp$ and finite intensity $\rho$,
 and let  the following
conditions be  satisfied:

\begin{itemize}

\item  (i)
For   some  constants $c_0>0 $ and $\a>-1$
\begin{equation}\label{cond_nu}
 0< \nu ( [-E,E ])\leq  c_0  |E|^{\a+1}  \,,\qquad \forall E\in (0,1]
 \,;
\end{equation}
\vspace{.1cm}


\item (ii)
There exist positive constants  $\rho', K  $ and there exists $p\in
(0,1]$  such that, setting
$$\Lambda _K(x)= x+[-K/2,K/2)^d $$
and defining the 
 random field $Y$   as 
\begin{equation}\label{serioso}
Y=\{Y(x)\,:\, x\in K \ZZ^d\}\,,\qquad 
 Y(x):=\hat \xi \left( \Lambda _K(x) \right)\,,
\end{equation}
then  the law of $Y$ when $\hat \xi$
is chosen   with law   $\hat \Pp^{(p)}$ (the $p$--thinning of $\hat
 \Pp$)
is stochastically dominated by the law of  $Y$ when 
 $\hat \xi$
 is chosen with  law $\hat \Pp_{\rho'}$ (the Poisson point process with density $\rho'$).
\end{itemize}

\noindent Then the $d\times d$ symmetric matrix $D(\b)$ solving the variational problem    (\ref{sereno})
 admits the following upper bound   for $\b$ large enough:
\begin{equation}
\label{eq-Mott} D(\b) \leq c_1 \b ^{\,c_2} \exp\left(-C
\,\beta^{\frac{\a+1}{\a+1+d} }\right)\II\, ,
\end{equation}for   suitable $\b$--independent positive constants $c_1, c_2 , C$.
\end{theo}

We recall that due to Strassen theorem  the stochastic domination
assumption in condition (ii) above is equivalent to the fact  that
one can construct on the same probability space  processes
$Y_1=\{Y_1(x)\,:\, x\in K \ZZ^d\}$ and $Y_2=\{Y_2(x)\,:\, x\in K \ZZ^d\}$  in
such a way that
\begin{equation}\label{accoppio}
Y_1(x) \leq Y_2 (x)\,, \qquad \forall x\in K \ZZ^d,
\end{equation}
  the law of $Y_1$ equals the law of  $Y$ when $\hat \xi$
is chosen   with law   $\hat \Pp^{(p)}$ and the law of $Y_2$
equals the law of  $Y$ when 
 $\hat \xi$
 is chosen with  law $\hat \Pp_{\rho'}$.

\smallskip

Theorem \ref{teo-Mottup} applies to the case that  $\hat \Pp$ is
or is dominated by a stationary Poisson point process $\hat \Pp _{\rho '}$, i.e.\ when  one can define random sets $(\xi,\xi')$
such that  $\xi\subset \xi'$ almost surely and  $\xi,\xi'$
have marginal distributions  $\hat \Pp$ and  $\hat \Pp _{\rho '}$ respectively. An example is given by 
  Gibbsian random point fields  with repulsive
interactions (cf. \cite{GK} and \cite{CF1}[Section 5]). 

Moreover, the above theorem
covers also
the case of thinnings  of point processes  with uniform bounds on the local density,  as for example  diluted  crystals. 
Note that in this case the point process  $\hat \Pp$  is not stochastically 
dominated by any  stationary Poisson point process. We refer to Section \ref{WS} for a more detailed discussion.

\smallskip

\subsection{Overview}   In Section \ref{basicfacts} we recall some definitions and
results about point processes  (see \cite{DVJ}, \cite{FSS}  for more
details) and state some technical results needed later on.
\smallskip

The proof of  Theorem  \ref{teo-Mottup} is based on the variational
formula (\ref{sereno}) since
 for each fixed  function $f\in L^\infty (\Pp _0 )$ the r.h.s. in
(\ref{sereno}) gives an upper bound on $(a,D(\b) a)$.    In Section \ref{sec_ricetta}   we
derive an upper bound on $D(\b)$  by taking   special   test functions $f$   in the r.h.s. of (\ref{sereno}).
 The choice of such test functions   is  inspired by \cite{PR} and is related to  the percolation approach
 of    \cite{AHL}, \cite{FSS}.  In Section \ref{poisson_case} we first show that   the above upper
 bound  together with some  scaling arguments leads to (\ref{eq-Mott}) if $\Pp_0$ is the $\nu$-randomization of 
 the Palm distribution  associated to   a stationary Poisson point
  process.  In Section  \ref{general_case}  we  extend the proof to  general point processes
  satisfying assumption (ii).
 
  As conjectured in the introduction of
  \cite{FSS}, the leading contribution to the conductivity  of the medium as $\b \uparrow \infty$  comes from
  a subset of impurities  that converges to
  a Poisson point process under suitable space rescaling. Due to the relevance of Poisson point
  processes in Mott's law and since in the Poissonian case the proof
  of Theorem \ref{teo-Mottup} is more transparent and simple, we
  have preferred to treat the Poissonian  case and the general case
  separately.

Finally, in Section \ref{WS} we show that Theorem \ref{teo-Mottup}
can be applied to  thinnings of point processes  with uniform bounds on the local density as diluted crystals.

For a
 more detailed discussion about  mathematical aspects and physical  motivations   of Mott random
 walk we refer to \cite{FSS},  \cite{SE}  and references therein.

\section{Simple point processes }\label{basicfacts}

In this section we recall some basic definitions and results about
simple point processes, referring to \cite{DVJ}, \cite{FSS} for more
details.

\smallskip

In what follows, given a topological set $Y$ we write $\Bb (Y)$ for
the $\sigma$--algebra of its Borel subsets.  We denote $\hat \Nn$
the space of simple counting measures $\hat \xi $ on $\RR^d$, i.e.\
integer--valued measures such that $\hat \xi (B)<\infty$ for all
bounded $B\in \Bb (\RR^d) $, and $\hat\xi (x)\in\{0,1\}$ for all $x\in\RR^d$.
 One can show that $\hat \xi \in \hat\Nn
$ if and only if $\hat \xi = \sum _j  \d_{x_j}$ where
$\{x_j\}\subset \RR^d $  is a locally finite set. 
  Trivially, a simple counting
measure $\hat\xi$  can be identified with its support. Given $x\in \RR^d$ the
translated counting measure $S_x\hat \xi$ is defined as
    $S_x \hat \xi = \sum _j  \d_{x_j-x} $ if $\hat \xi=\sum _j  \d_{x_j}$.

 The space  $\hat \Nn$ is endowed with  the $\sigma$--algebra of measurable subsets generated by the maps
 $$
 \hat\Nn\ni\hat  \xi \rightarrow \hat \xi (B)\in \NN \,, \qquad  B\in \Bb (\RR^d)  \text{ bounded}\,.
 $$ A {\sl simple  point process} (on $\RR^d$) is a measurable
 map $\Phi$ from a probability space into $\hat\Nn$. 
With abuse of notation, we  identify a simple 
point process with its distribution  $\hat \Pp $  on $\hat \Nn $.
Moreover,  one  calls it   {\sl stationary}
 if $\hat \Pp (A)= \hat \Pp (S_x A)  $, for all $x\in \RR^d$ and $A\subset \hat \Nn $ measurable.
In this case, we define the $\k$--moment $\rho_\k$ as  
\begin{equation}
\rho _\k:= \EE _{\hat \Pp } \left( \hat \xi \bigl( [0,1]^d \bigr)
^\k \right)\,, \qquad \k >0\,.
\end{equation}
Then $\rho =\rho_1$ is the so--called {\sl intensity} of the process.

\smallskip

The {\sl Palm distribution} $\hat \Pp _0  $  associated to  a
stationary simple point process $\hat \Pp $ on $\RR^d$ with finite
intensity $\rho$  is the probability measure  on the measurable
subset $\hat\Nn _0\subset \hat \Nn $,
$$
\hat \Nn _0 :=
\left\{\hat\xi \in \hat \Nn \,:\, \hat \xi (0)= 1\right \}\, ,
$$
characterized  by
the  Campbell identity:
\begin{equation}\label{campbell}
\hat \Pp_0 ( A)  =
\frac{1}{\rho K^d}   \int_{\hat \Nn}  {\hat \Pp} ( d \hat \xi) \int_{Q_K} \hat{\xi}
(d x) \, \chi_A ( S_x \hat \xi),\qquad  \forall  A\subset \hat \Nn _0 \text{ measurable}\,,
\end{equation}
where $Q_K = [-K/2, K/2]^d$, $K>0$ (since $\hat\Pp$ is stationary,
the r.h.s. in Campbell identity does not depend on $K$). As
discussed in \cite{DVJ}, the point process $\hat \Pp_0$ can be
thought of as
 obtained from the
 point process $\hat \Pp$ by conditioning  the latter to give positive mass at the origin.

\smallskip 
Given two
 simple point processes $\hat \Pp$, $\hat \Pp '$ one says that $\hat \Pp$
is {\sl stochastically dominated} by  $\hat \Pp '$, shortly  $\hat \Pp
\preceq \hat \Pp '$, if 
there exists  a coupling of
$ \hat \Pp, \hat \Pp ' $ such that almost surely $\hat \xi \subset\hat \xi' $, with $(\hat\xi,\hat \xi')$ denoting the random sets with marginal
distributions given by  $\hat \Pp$ and  $\hat \Pp '$  respectively.
 We refer to
\cite{GK} for more details on stochastic domination between  point processes.

\smallskip

We denote $ \Nn$ the space of  (marked) simple counting measures $  \xi $
on $\RR^d\times[-1,1]$ such that $\xi (B\times [-1,1])<\infty$ for
all  $B\in \Bb (\RR^d)$ bounded, and $\xi( \{x\}\times [-1,1] )\in\{0,1\}$ for all $x\in\RR$. 
One can show that $ \xi \in  \Nn $
if and only if $\xi = \sum _j \d_{(x_j, E_j) }$, where $\{(x_j,
E_j) \}\subset \RR^d\times [-1,1] $  is a locally finite set such that for each $x\in\RR$ there is
at most one couple $(x_j,E_j)$ with $x_j=x$.  Trivially, a simple counting measure $\xi$ can be identified
with its support. The value $E_j$ is called the {\sl  mark}   at
$x_j$. For physical reasons, we call it the {\sl energy mark}.
 Given
$\xi \in\Nn$ we write $\hat \xi$ for the simple counting measure on $\RR^d$
  defined as $\hat \xi (B)=
\xi (B\times[-1,1] )$, for all $B\in \Bb (\RR^d)$ bounded. 
  Given $x\in \RR^d$ the translated simple  counting measure $S_x  \xi$ is defined as
    $S_x \xi = \sum _j  \d_{(x_j-x, E_j)} $ if $\xi=\sum _j  \d_{(x_j,E_j)}$.

\smallskip

 The space  $  \Nn$ is endowed with  the $\sigma$--algebra of measurable subsets generated by the maps
 $$
 \Nn\ni \xi \rightarrow \xi (B)\in \NN\,,\qquad  B\in \Bb (\RR^d\times [-1,1]) \text{ bounded}.
 $$
 A {\sl marked simple   point process on $\RR^d$} is a measurable map $\Phi$
 from a probability space into $\Nn$. Again, with abuse of notation,
 we will identify it  with its distribution  $ \Pp $
   on $ \Nn $. Moreover, it  is called  {\sl stationary}
 if $\Pp (A)=  \Pp (S_x A)  $, for all $x\in \RR^d$ and $A\subset   \Nn $ measurable.
In this case,  we define the $\k$--moment $\rho_\k$ as
\begin{equation*}
\rho _\k:= \EE _{ \Pp } \left( \hat \xi\bigl( [0,1]^d \bigr
)^\k\right)\,, \qquad \k >0\,,
\end{equation*}
and  call $\rho:=\rho_1 $ the intensity of the process.

\smallskip

The {\sl Palm distribution} $\Pp_0 $  associated to  a stationary marked simple  point process $\Pp $
  with finite intensity $\rho$  is the probability measure on the
measurable subset $ \Nn _0\subset   \Nn $,
$$
  \Nn _0 :=
\left\{ \xi \in  \Nn \,:\, \hat \xi (0)= 1\right \}\, ,
$$
 \smallskip
characterized by
the   Campbell identity
\begin{equation}\label{campbellbis}
 \Pp_0 ( A)  =
\frac{1}{\rho K^d}   \int_{\Nn}  {\Pp} ( d \xi) \int_{Q_K} \hat{\xi}
(d x) \, \chi_A ( S_x \xi),\qquad  \forall  A\subset   \Nn _0 \text{
measurable}\,, K>0\,.
\end{equation}

\bigskip

A standard procedure for obtaining a marked simple  point process
from a given simple point process on $\RR^d$ is the {\sl
$\nu$--randomization}, where $\nu$ is a probability measure on $[-1,
1]$:
  given a  realization of the   simple point process on $\RR^d$, its points are marked by
 i.i.d. random variables with common law $\nu$. It is simple
   to check that the
  $\nu$-randomization of the Palm distribution
   associated to a given stationary 
  simple point process coincides with the Palm distribution associated to the
   $\nu$--randomization of the stationary  simple point process.

\bigskip



We conclude this section recalling some technical results derived in
\cite{FSS}. In particular, point (i) of the following lemma follows
from \cite{FSS}[Lemma 1, (i)] and the Monotone Convergence Theorem,
 while the proof of point (ii)  is similar to the proof of \cite{FSS}[Lemma 2]:
 \begin{lemma}\label{techn_lemma} \cite{FSS}  Let $\Pp _0$ be the Palm distribution
  associated to a stationary marked simple   point process. \\
\noindent {\rm (i)}
Let  $f:\Nn_0\times\Nn_0 \to \RR$ be a measurable  function which is non negative
 or    such that
$\int \hat{\xi} ( dx) \,|f (\xi, S_x \xi )|$ and
$\int \hat{\xi} ( dx) \,|f ( S_x \xi, \xi )|$
are in $L^1(\Pp_0)$.
Then
\begin{equation*}
\int \Pp_0(d\xi) \int \hat{\xi} ( dx)   f(\xi,S_x \xi)
=
\int \Pp_0(d\xi) \int \hat{\xi} ( dx)   f(S_x \xi, \xi) \mbox{ . }
\end{equation*}

\vspace{.1cm}

\noindent {\rm (ii)} Let $n$ be a nonnegative integer such that
$\rho _{n+1 } <\infty$. Then
$$
\int \Pp_0(d\xi) \left( \int \hat{\xi} ( dx)   e^{-\g |x|} \right)^n
<\infty
$$
for any $\g>0$.

\end{lemma}

%
%
\section{Upper bounds via special test functions}\label{sec_ricetta}

In this section we let $\Pp_0$ be the Palm distribution associated
to the $\nu$--randomization of a  stationary simple point process
with  finite intensity and obtain upper bounds on $D(\b)$ by
choosing special test functions $f\in L^\infty(\Pp _0)$ in the
r.h.s. of (\ref{sereno}).   We suppose that $\nu $ is not
concentrated in a unique value, i.e. $\nu$ is not of the form $\nu =
\d_E$. This implies that  for $\Pp _0$--a.a. $\xi$, $S_x \xi \not =
S_y \xi$ if $x,y\in \hat \xi$ and $x\not= y$. Note that  the above
assumption is satisfied  whenever (\ref{cond_nu}) is fulfilled, moreover in the case 
$\nu=\d_E$ the $\b$--dependence  of the diffusion matrix  is trivial.

\smallskip

Given $\xi \in \Nn _0$, let
 $\Ee^\b (\xi)$  be a family of non oriented links
in $\hat \xi$, i.e.
\begin{equation}\label{tamburo}
\Ee^\b(\xi)\subset \left\{
\,\{x,y\}\,:\, x,y\in \hat \xi\text{ and } x\not =y\,\right\},
\end{equation}
covariant w.r.t. space translations, i.e.
\begin{equation}\label{covariante}
 \Ee^\b(S_x \xi)=\Ee^\b(\xi)-x,\qquad \forall \xi\in \Nn_0, \,x\in \hat \xi\, .
\end{equation}
Consider the graph $\Gg^\b(\xi)$ with vertexes set $\Vv^\b(\xi)$ and
edges set $\Ee^\b(\xi)$ where
\begin{equation}\label{chitarra}
\Vv^\b(\xi)= \left\{x\in \hat \xi \,:\,
\exists y \in \hat\xi \text{ with } \{x,y\}\in \Ee^\b(\xi)\right\}.
\end{equation}
Given $x,y$ in $\Vv^\b(\xi)$ we say that they are  connected if
there exists a path   in $\Gg^\b(\xi)$ going from $x$ to $y$.
Moreover,
 we denote by $C_x^\b (\xi)$   the connected component in
$\Gg^\b (\xi)$ containing $x$  if $x\in \Vv ^\b (\xi)$, while we set  $ C_x^\b (\xi)=\emptyset$ if $x\in \hat \xi \setminus  \Vv ^\b
(\xi)$.

\begin{prop}\label{ricetta} Let $\Pp_0$ be the Palm distribution associated to the
 $\nu$--randomization of a stationary simple point process with $\rho _2<\infty$ and  
 $\nu\not =\d_E$ for any $E\in [-1,1]$.
Suppose that for each $\b>0$ a random  graph $\Gg^\b(\xi)= \left(
\Vv ^\b (\xi), \Ee ^\b (\xi) \right)$,
     satisfying (\ref{tamburo}), (\ref{covariante}) and (\ref{chitarra}),       is assigned and that the
 following assumptions (A1), (A2) are  fulfilled:
\begin{itemize}
 \item (A1)
 There exists a positive function $\ell (\b)$ such that
 \begin{equation}
 |x-y|>\ell (\b) \Rightarrow \{x,y\}\not\in \Ee^\b(\xi) ,\qquad \forall x,y \in
 \hat \xi\, , \;\forall \xi \in \Nn_0 \, ;
 \end{equation}
 \item
(A2) The function $\Nn_0\ni \xi \rightarrow C_0^\b (\xi) \in \Nn$ is
measurable and for some $\e>0$
\begin{align}
&
\limsup _{\b \uparrow \infty}
\EE _{\Pp_0} \left(|
C_0^\b (\xi)|^{2+\e}\right)<\infty\, , \label{viola}
  \\
&
 \rho _{\lceil 2(1+\e)/\e \rceil+1}<\infty\, ,\label{iris}
\end{align}
\end{itemize}
where $\lceil a \rceil$ denotes the smallest integer larger than
$a$.

 Then, for all  $i=1,\dots, d$  and $\b$ large enough,
\begin{equation}\label{everest}
D_{i,i}(\b )\leq   6 \int \Pp_0(\xi) \int \hat{\xi} (dx) c_{0,x}
(\xi) \left( \bigl(x^{(i)}\bigr) ^2+ \ell(\b)^2 |C_0^\b (\xi)|^2
\right)
 \II_{  x\not \in C_0^\b(\xi)}\,,
\end{equation}
where $x^{(i)}$ denotes the $i$--th coordinate of $x$   and $
\II_{ x\not \in C_0^\b(\xi)}$ is the characteristic function of the
event $\{x\not \in C_0^\b(\xi)\}$.

Moreover, suppose that the following additional assumption is fulfilled:
\begin{itemize}
 \item (A3) There exists a positive function $C(\b)$ such that
 \begin{equation}
 \{x,y\}\not \in \Ee ^\b (\xi) \Rightarrow c_{x,y} (\xi)\leq C(\b),\qquad \forall x,y\in \hat
 \xi\,, \;   \forall \xi \in \Nn_0\,.
 \end{equation}
\end{itemize}
Then, for all  $i=1,\dots, d$, $\k \in (0,1)$ and $\b$ large enough,
\begin{equation}\label{k2}
   D_{i,i}(\b )\leq    c(\k )\,  C(\b)^\k (1+ \ell (\b)^2)\,,
\end{equation}
for a suitable positive constant $c(\k )$ depending on
$\k $, but not on $\b$. In particular,
\begin{equation}\label{montebianco}
D(\b)\leq d \, c(\k)\,  C(\b)^\k (1+ \ell(\b)^2)\II \,.
\end{equation}
\end{prop}

 The proof of the above Proposition is
obtained by plugging  suitable test functions $f$ in the variational
formula (\ref{sereno}). Our test functions are similar to the ones
used in \cite{PR}[Proof of Theorem 3.12].

\begin{rem}\label{fine}
As one can easily deduce  from the proof of the above  
Proposition, the results  (\ref{everest}), (\ref{k2}) and
(\ref{montebianco}) hold for all $\b>0$ if condition (\ref{viola})
is  satisfied with $\limsup_{\b\uparrow \infty}$ replaced by
$\sup_{\b>0}$.
\end{rem}
%
%
\begin{proof} Assume (A1), (A2) to be satisfied and,
 given a positive integer $N$,   consider the test function $f^\b
_N:\Nn_0 \rightarrow \RR_{\geq 0} $ defined as follows:
\begin{equation}
f_N^\b (\xi)=
\begin{cases}
-\min \left\{ z^{(i)}\,:\, z\in C_0 ^\b (\xi)\right\}
 & \text{ if } 1\leq
|C_0^\b
(\xi)|\leq N,\\
0 & \text{ otherwise},
\end{cases}
\end{equation}
where $z^{(i)}$ denotes the $i$--th coordinate of $z$.
Due to (A2)  $f^\b_N$ is measurable,  while due to (A1)
\begin{equation}
0\leq f^\b _N(\xi)\leq |C_0^\b (\xi)| \ell (\b)\,,\qquad \forall
\xi\in \Nn_0\, .
\end{equation}
In particular,
\begin{equation}\label{roma}
  \left( x^{(i)}  -\nabla_x f_N^\b (\xi) \right)^2 \leq
3\left( \bigl (  x^{(i)}\bigr)  ^2+ \ell (\b)^2 |C_0^\b (\xi)|^2+
\ell (\b)^2 |C_0^\b (S_x\xi)|^2\right)\,, \qquad \forall \xi
\in\Nn_0\, .
\end{equation}
 Due to (\ref{sereno})
\begin{equation}\label{passo0}
D_{i,i}(\b ) \leq
\int \Pp_0(\xi) \int \hat{\xi} (dx) c_{0,x}  (\xi)
 \left( x^{(i)} -\nabla_x f_N^\b (\xi) \right)^2 = I^{(1)}_N(\b)+I^{(2)}_N(\b)+I^{(3)}_N (\b) \, ,
\end{equation}
where
\begin{align*}
& I^{(1)}_N(\b)=\int \Pp_0(\xi) \int \hat{\xi} (dx) c_{0,x}  (\xi)
\left( x^{(i)} -\nabla_x f_N^\b (\xi) \right)^2 \II_{ x\in
C_0^\b (\xi)}\II _{|C_0^\b (\xi)|\leq N}\, ,
\\
& I^{(2)}_N (\b)=\int \Pp_0(\xi) \int \hat{\xi} (dx) c_{0,x}  (\xi)
\left( x^{(i)} -\nabla_x f_N^\b (\xi) \right)^2 \II_{ x\in
C_0^\b (\xi)}\II _{|C_0^\b (\xi)|> N}\,,
\\
 &I^{(3)}_N (\b)=\int \Pp_0(\xi) \int \hat{\xi} (dx) c_{0,x}  (\xi) \left( x^{(i)}
 -\nabla_x f_N^\b (\xi) \right)^2 \II_{ x\not \in  C_0^\b
(\xi)}\, ,
\end{align*}
where $\II_A$ denotes that characteristic function of the event $A$.

\smallskip

\noindent $\bullet$ {\em Estimate of $I^{(1)}_N (\b)$}. If  $x\in
C_0^\b(\xi)$ and $|C_0^\b (\xi)|\leq N$, then
 \begin{equation}\label{traslo}
C_0^\b (S_x \xi) = C_0^\b (\xi) -x \end{equation} and in particular
\begin{equation}\label{traslo1}
| C_0^\b (S_x \xi)|=|C_0^\b (\xi)|\in [1,N]\, .
\end{equation}
Due to (\ref{traslo}) and (\ref{traslo1}) we get $f_N^\b(S_x\xi)=
x^{(i)} + f_N^\b (\xi)$ and therefore
  $x^{(i)} -\nabla _x f_N^\b (\xi)=0$, thus implying
 \begin{equation}\label{passo1}
 I^{(1)}_N(\b) =0\, .
\end{equation}

\noindent $\bullet$ {\em Estimate of $I^{(2)}_N (\b) $}. We claim
that, if $\b$ is large enough,
\begin{equation}\label{passo2}
\lim _{N\uparrow \infty} I^{(2)}_N (\b ) =0\,.
\end{equation}
In order to prove the above limit we observe that,
 due to
(\ref{roma}) and since $c_{0,x}(\xi)\leq e^{-|x|}$,
\begin{equation}
I^{(2)}_N (\b)\leq  3 \left(J^{(1)}_N(\b)+\ell (\b)^2  J^{(2)}_N
(\b)+\ell (\b) ^2  J^{(3)}_N (\b)\right)\,,
\end{equation}
where
\begin{align*}
& J^{(1)}_N (\b)=\int \Pp_0(\xi) \int \hat{\xi} (dx) e^{-  |x|}|x|
^2 \II _{|C_0^\b (\xi)|> N}\,,
\\
&  J^{(2)}_N(\b)=\int \Pp_0(\xi) \int \hat{\xi} (dx)  e^{-
|x|}|C_0^\b (\xi)|^2  \II_{ x\in C_0^\b (\xi)}\II _{|C_0^\b (\xi)|>
N}\, ,
\\
& J^{(3)}_N (\b) =\int \Pp_0(\xi) \int \hat{\xi} (dx) e^{- |x|}
|C_0^\b (S_x \xi)|^2  \II_{ x\in C_0^\b (\xi)}\II _{|C_0^\b (\xi)|>
N}\, .
\end{align*}
Due to   (A2), $\lim _{N\uparrow \infty} \II _{|C_0^\b
(\xi)|> N}=0$ for $\Pp_0$--a.a. $\xi$. Hence, by the Dominated
Convergence Theorem, we only need to prove that
\begin{align}
&  \int \Pp_0(\xi) \int \hat{\xi} (dx) e^{-  |x|}|x|^2
<\infty\, ,\label{mosca1}
\\
&    \int \Pp_0(\xi) \int \hat{\xi} (dx)  e^{- |x|}|C_0^\b (\xi)|^2
\II_{ x\in C_0^\b (\xi)} <\infty\, ,\label{mosca2}
\\
&   \int \Pp_0(\xi) \int \hat{\xi} (dx) e^{- |x|} |C_0^\b (S_x
\xi)|^2 \II_{ x\in C_0^\b (\xi)} <\infty\, ,\label{mosca3}
\end{align}%
 for $\b$ large enough.

Since $\rho_2<\infty$,  Lemma \ref{techn_lemma} (ii)  implies
(\ref{mosca1}). Since
$$ \text{l.h.s. of (\ref{mosca2}) } \leq
  \int \Pp_0(\xi) \int \hat{\xi} (dx)  e^{- |x|}|C_0^\b (\xi)|^2
  \,,
     $$
given $p,q>1$ with $1/p+1/q=1$, due to H\"older inequality
$$
\text{l.h.s. of }(\ref{mosca2}) \leq
    \EE_{\Pp_0} \left( \,
   |C_0^\b (\xi)|^{2q} \right)^{1/q}
\EE_{\Pp_0} \left( \, \left(
 \int \hat{\xi} (dx)  e^{-  |x|} \right)^p\,\right)^{1/p}
.
$$
Choosing  $2q=2+\e$ (hence $p=(2+\e)/\e $) we have that, due to
(\ref{viola}), (\ref{iris}) and  Lemma \ref{techn_lemma} (ii),
both the factors in the r.h.s. are finite   for $\b$ large enough.
 Hence (\ref{mosca2}) is true.  Finally we observe
that  the    l.h.s. of (\ref{mosca2}) equals
the l.h.s. of  (\ref{mosca3}) due to
Lemma \ref{techn_lemma} (i). In fact, consider the function $f$
defined on $ \Nn_0\times \Nn _0$ as
$$
f(\xi, \zeta)=
\begin{cases}
e^{-|x|} |C_0^\b (\xi)|^2  \II_{ x\in C_0^\b (\xi)}   &
\text{ if }\zeta=S_x \xi,\; x\in \hat \xi \,, \\
 0 & \text{ otherwise}\, .
\end{cases}
$$
Note that the above definition is well posed $\Pp_0$--a.s. since due
to the choice  $\nu\not=\d_E$ we have that  $S_x \xi \not=
S_y\xi $ if $x,y\in \hat \xi$, $x\not = y$,   for $\Pp_0$--a.a.
$\xi$. Then, if $x\in \hat \xi$,
 \begin{align*} &  f(\xi, S_x\xi)= e^{-|x|} |C_0^\b (\xi)|^2 \II_{ x\in C_0^\b (\xi)}  \, ,\\
&  f(S_x\xi , \xi)= e^{-|-x|}    |C_0^\b (S_x \xi)|^2 \II_{ -x\in
C_0^\b (S_x \xi)} = e^{-|x|}    |C_0^\b ( \xi)|^2 \II_{ x\in C_0^\b
( \xi)} \,.
 \end{align*}
 Note that in the
last identity we have used (\ref{covariante}) which  implies that
$$
|C_0^\b (S_x \xi)|^2 \II_{ -x\in C_0^\b (S_x \xi)} = |C_0^\b (
\xi)|^2 \II_{ x\in C_0^\b ( \xi)} . $$ Hence, due to Lemma
\ref{techn_lemma} (i) we conclude that the l.h.s. of (\ref{mosca2})
equals the l.h.s. of (\ref{mosca3}).
\smallskip

\noindent $\bullet$ {\em Estimate of $I^{(3)}_N(\b) $}. Due to
(\ref{roma}) we can bound
\begin{equation}\label{passo3}   
I^{(3)}_N(\b)\leq 3 \int \Pp_0(\xi) \int \hat{\xi} (dx) c_{0,x}
(\xi) \left( \bigl(x^{(i)} \bigr)^2 + \ell (\b)^2 |C_0^\b (\xi)|^2+
\ell (\b)^2 |C_0^\b (S_x\xi)|^2 \right)
 \II_{ x\not \in  C_0^\b(\xi)}\,.
\end{equation}
By defining   now $f(\xi,\zeta)$ as
$$
f(\xi, \zeta)=
\begin{cases}
c_{0,x}(\xi) |C_0^\b (\xi)|^2  \II_{ x\not \in C_0^\b (\xi)}   &
\text{ if }\zeta=S_x \xi,\; x\in \hat \xi \,, \\
 0 & \text{ otherwise}\, 
\end{cases}
$$
and  reasoning as in the proof that the l.h.s. of (\ref{mosca2})
equals the l.h.s. of (\ref{mosca3}),
 it is simple to show that
\begin{equation*}\label{simili}
\int \Pp_0(\xi) \int \hat{\xi} (dx) c_{0,x}  (\xi)  |C_0^\b (\xi)|^2
\II_{ x\not \in  C_0^\b(\xi)} = \int \Pp_0(\xi) \int \hat{\xi} (dx)
c_{0,x} (\xi)  |C_0^\b (S_x \xi)|^2 \II_{x\not \in C_0^\b(\xi)}\, .
\end{equation*}


Hence
\begin{equation}\label{bianchetto}
I^{(3)}_N(\b)\leq 6 \int \Pp_0(\xi) \int \hat{\xi} (dx) c_{0,x}
(\xi) \left( \bigl(x^{(i)} \bigr)^2+ \ell (\b)^2 |C_0^\b (\xi)|^2
\right)
 \II_{ x\not \in  C_0^\b(\xi)}\,.
\end{equation}

\bigskip

$\bullet$ {\em Conclusions.} The bound (\ref{everest}) follows from
(\ref{passo0}), (\ref{passo1}), (\ref{passo2}) and
(\ref{bianchetto}). Suppose now that also  assumption (A3) is valid.
In particular,
 if   $ x\not\in  C_0^\b (\xi)$  then $\{0,x\}\not \in
\Ee^\b (\xi)$ and therefore $c_{0,x}  (\xi) \leq C(\b)$. In
particular $c_{0,x}(\xi)\leq C(\b)^\k e^{-(1-\k)|x|}$, for all
$\k\in (0,1)$.   Due to (\ref{everest}) we get
\begin{multline}\label{pietra}
  D_{i,i}(\b, \g)\leq   6\, C(\b)^{\k}  \int \Pp_0(\xi)
\int \hat{\xi} (dx) e^{-(1-\k)  |x|} \left( |x|^2+ \ell (\b)^2
|C_0^\b (\xi)|^2
\right)\\
\leq  C(\b)^{\k}  c( \k)(1+  \ell (\b)^2),
 \end{multline}
where the last bound follows from (\ref{viola}) and  the same
arguments used for proving (\ref{mosca2}). Hence the proof of  
(\ref{k2}) is concluded.

Finally, let us prove  (\ref{montebianco}).  The matrix $D(\b )$ is positive
and symmetric. In particular,
 $$(a, D(\b ) a)\leq \max \{\lambda_i\,:\, 1\leq i\leq d\}(a,a)\,, \qquad \forall a\in \RR^d,
$$
  where
$\lambda_1,\dots,\lambda_d$ are the eigenvalues of $D(\b)$. Since
$\lambda_i\geq 0$ for each $i$,
$$
(a, D(\b ) a) \leq \Tr \left( D(\b )\right) (a,a) = \left(\sum
_{i=1}^d D_{i,i}(\b ) \right) (a,a)\,, \qquad \forall a\in \RR^d,
$$
and   (\ref{montebianco}) follows from  (\ref{k2}).
\end{proof}


\section{Proof of Theorem \ref{teo-Mottup} in the Poissonian case}\label{poisson_case}


In this section   we prove Theorem \ref{teo-Mottup} in  the special
case that $\Pp_0$ is the
 Palm distribution  associated to the the $\nu$--randomization of
 the
  Poisson point process with intensity $\rho>0$. In the next section,
   we extend the proof to the general case.
We write $\hat \Pp _\rho$ for the Poisson point process
 with density $\rho$, $\Pp _\rho$ for its $\nu$--randomization,
$\hat \Pp _{0,\rho}$ for the Palm distribution associated to $\hat
\Pp_\rho$ and $ \Pp_{0,\rho}$ for the Palm distribution associated
to $ \Pp _\rho$. Equivalently, $ \Pp_{0,\rho}$ is the
$\nu$--randomization of $\hat \Pp_{0,\rho}$. In what follows, we
write $Q_L$ for the cube $[-L/2,L/2]^d$.


\smallskip The proof is based on   Proposition \ref{ricetta},
 scaling arguments and   continuum percolation. We recall some results of continuum
 percolation referring to \cite{MR} for a more detailed  discussion.
Given $r>0$ we write  $B_r (x)$ for the closed ball centered at
$x\in \RR^d $ with radius $r$.  If $x=0$ we simply write $B_r$.
We define the {\sl occupied region}  of the Boolean model with
radius $r$ driven by $\hat\xi\in \hat\Nn $ as
 $$
\XX_r (\hat\xi):=\cup_{x\in \hat \xi} B_r(x )=\{y\in\RR^d\,:\, d(y, \hat \xi )\leq r\}\,,
$$
where $d(\cdot,\cdot)$ denotes the euclidean distance.  The connected components in the occupied region will be called {\sl occupied components}. For $A\subset \RR^d$, we denote by $W_r (A)=W_r(A)[\hat\xi]$  the union of all occupied components having  non--empty
 intersection with $A$.
  Given $A\subset \RR^d$ and $B\subset \RR^d$ we write $A\stackrel{r}\longleftrightarrow  B$
  if there exists a path inside  $ \XX_r (\hat\xi) $ connecting $A$ and $B$.


\smallskip

The following results hold for stationary Poisson point processes
\cite{MR}: there exists a positive density  $\rho_c$ such that
for all bounded  subsets  $A\subset \RR^d, \, A\not =  \emptyset $,
\begin{align}
& \rho _c =\inf  \left\{ \rho>0\,:\, \hat \Pp _\rho \left[ \text{ diam } W_1 (A)=\infty \right]>0\right\},\\
&
\hat \Pp _\rho \left( A \stackrel{1}\longleftrightarrow  \partial Q _L  \right) \leq e^{-c(\rho, A) \, L}, \qquad \forall L>0,\qquad \forall \rho <\rho_c
\end{align}
for a suitable positive constant $ c(\rho,A) $ depending on $\rho,A
$. Above $\partial Q_L$ denotes the border of the cube $Q_L$.
Moreover,  one can prove that all occupied components in $\XX_1$ are bounded $\hat \Pp_{\rho}$--a.s. for $\rho<\rho_c$, while there exists a unique unbounded occupied  component
 in $\XX_1$    $\hat \Pp_{\rho}$--a.s. for $\rho>\rho_c$.

\smallskip

Note that the function $\RR^d \ni x\rightarrow x/r \in \RR^d$ maps $\hat \Pp _\rho$ in
$\hat\Pp _{\rho \,r^d}$,
  namely  if $\hat \xi$ has
  law $\hat\Pp _{\rho}$ then $\{x/r\,:\, x\in \hat\xi\}$
  has law $\hat \Pp _{\rho\, r^d }$. This  scaling property allows to restate
  the above results for a fixed density $\rho$ and
  varying radius $r$:
     the positive constant
\begin{equation}\label{rcritico}
r_c(\rho):= (\rho _c/\rho)^{1/d}= \rho ^{-1/d} r_c (1)
\end{equation}
satisfies
\begin{align}
&  r_c(\rho)= \inf
  \left\{ r>0 \,:\, \hat \Pp _\rho \left[ \text{ diam } W_r (A)=\infty \right]>0\right\},\\
&  \label{carrere}
\hat \Pp _\rho \left( A  \stackrel{r}\longleftrightarrow  \partial Q _L \right) \leq e^{-c(r,\rho,A) \, L}, \qquad \forall L>0,\qquad \forall r<r_c(\rho)  ,
\end{align}
 for all bounded subsets  $A\subset \RR^d, \, A\not =  \emptyset $,
  and for a suitable positive constant $c(r,\rho,A) $ depending on $r,
\rho, A$.
Moreover,  all occupied components
in $\XX_r$ are bounded $\hat \Pp_{\rho}$--a.s. for $r<r _c$,
while there exists a unique unbounded occupied  component
 in $\XX_r$    $\hat \Pp_{\rho}$--a.s. for $r>r_c$.

We point out a simple consequence of (\ref{carrere}):
\begin{lemma}\label{boundedmom}
If $r<r_c (\rho)$ then for all bounded  sets  $A\subset \RR^d$ with $A\not = \emptyset$ and for all
$s>0$
\begin{equation}
 \EE_{\hat \Pp _\rho} \left( \hat \xi \left( W_r(A) [\hat \xi ] \right)^s \right)  <\infty\, .
 \end{equation}
\end{lemma}
\begin{proof}
Due to the stationarity of $\hat \Pp_\rho$, we can assume that $0\in
A$ without loss of generality.
 Fixed $p,q>1$ with $1/p+1/q=1$, by H\"older inequality and (\ref{carrere}) we get
\begin{multline}\label{paganini}
  \EE_{\hat \Pp _\rho} \left( \hat \xi \left( W_r(A)[\hat\xi]\right )^s \right) \leq \sum _{L=1}^\infty
   \EE_{\hat \Pp _\rho} \left( \hat\xi (Q_L )^s \, \II _{\{  A     \stackrel{r}\longleftrightarrow \partial Q_{L-1}  \}  }\right) \\
    \leq   \sum _{L=1}^\infty
   \EE_{\hat \Pp _\rho} \left( \hat\xi (Q_L )^{s\,p}\right)^{1/p}  \hat\Pp_{\rho}\left(  A     \stackrel{r}\longleftrightarrow \partial Q_{L-1}    \right) ^{1/q}\leq
    \sum _{L=1}^\infty
   \EE_{\hat \Pp _\rho} \left( \hat\xi (Q_L )^{s\,p}\right)^{1/p}  e^{-cL}.
\end{multline}
The thesis then follows by observing that, since $ \hat\xi (Q_L ) $ is a Poisson random variable with expectation $\rho L^d$,
$$
  \EE_{\hat \Pp _\rho} \left( \hat\xi (Q_L )^n \right) \leq c(\rho ) L^{d\,n} \,,\qquad \forall L>0,\;n\in \NN\,,
  $$
  thus implying that the last member in (\ref{paganini}) is summable.
   \end{proof}


We can now give the proof of Theorem \ref{teo-Mottup} for Poisson
point processes:

\smallskip

\noindent {\em Proof of Theorem \ref{teo-Mottup} for $\Pp _0= \Pp
_{0,\rho}$}.
 Given $\xi \in \Nn _0 $ we set
\begin{align*}
& E(\b): = \b ^{ -\frac{d}{\a +1+d } }\, ,\\
& \rho (\b ):= \rho \, \nu \left([-E(\b),E(\b)]\right)\,,\\
& \ell (\b):=   r_c\left(  \rho(\b) \right)  =  \rho(\b) ^{-1/d} r_c (1) \,,\\
&   \Ee ^\b (\xi):= \left\{
 \{x,y\}\,:\, x,y \in \hat \xi,\, x\not =y,\, |E_x|\leq E(\b),\, |E_y|\leq E(\b)\,   \text{ and } |x-y|\leq \ell (\b) \right\}\, .
\end{align*}
 We point out that  
  we could have defined  $\ell (\b)= \g\,
r_c\left(  \rho(\b) \right)$ for an
 arbitrary $\g\in (0,2)$. Here $\g:=1$. 
Assumption (\ref{cond_nu}) implies that
\begin{equation}
0<\rho(\b)  \leq   c_0 \,\rho \,   \b ^{ -\frac{d(\a+1)}{\a +1+d } }\,,
\end{equation}
hence
\begin{equation}
\label{rameau}
 \ell (\b) \geq  (c_0 \,\rho)^{-1/d}  r_c(1)\,
 \b ^{ \frac{\a+1}{\a +1+d } }\,.
\end{equation}
In particular, due to (\ref{uuu}),
\begin{multline}
\label{debussy} \{x,y\}\not\in \Ee^\b (\xi) \\ \Rightarrow
c_{x,y}(\xi)\leq \exp\left\{-\ell(\b) \wedge [\k_1 \b E(\b)]\right
\}\leq \exp \left\{
 -c(\a,\rho  )
 \b ^{   \frac{\a+1 }{\a +1+d }   }   \right \}=:C(\b)\,.
\end{multline}
Then, due to (\ref{montebianco}), in order to conclude the proof of
Theorem \ref{teo-Mottup} it is enough to check that   the
assumptions of Proposition \ref{ricetta} are fulfilled  when
the graph $\Gg ^\b(\xi) = \bigl( \Vv ^\b (\xi), \Ee^\b (\xi) \bigr)$
is defined via (\ref{chitarra}).   Conditions (\ref{tamburo}),
(\ref{covariante}) and  (\ref{chitarra}) are trivially satisfied.
(A1) is obvious, (A3) has already been checked: it therefore only
remains to consider (A2). Since Poisson point processes have finite
moments, the non trivial condition to be checked  is given by
(\ref{viola}), which can be justified by means of scaling arguments
and Lemma \ref{boundedmom} as follows.

\smallskip
 As discussed in \cite{DVJ},
  the process
$\hat \xi $ with law $\hat \Pp _{0,\rho}$ can be constructed by setting $\hat \xi: = \hat \omega \cup \{0\}$,
where   $\hat \omega$ is  a Poisson point process with law
 $\hat \Pp  _\rho$. Let $\omega$ be the $\nu$--randomization of the process $\hat \omega$ and let $E_0$ be a random variable with law $\nu$, independent from $\omega$. Then  $\xi:=\omega\cup\{ (0, E_0) \}$ has law $\Pp _{0,\rho} $.
  Setting
$$
\hat \omega _\b =\left \{ x\in \hat \omega\,:\, |E_x|\leq E(\b) \right\}\, , \qquad r(\b)=\ell (\b)/2\, ,
 $$
 we get
 \begin{equation}\label{musica}
  |C_0 ^\b (\xi)| \leq 1+ \hat \omega  _\b \left( W_{r(\b)} \left( B_{r(\b)}\right )[ \hat \omega _\b]  \right)\,.
  \end{equation}
Note that the process $\hat \omega _\b$, obtained  by thinning the
Poisson process $\hat\omega$ with density $\rho$,    has law $\hat
\Pp _{\rho (\b)}$. We now  consider  the space rescaling
 $$\RR^d\ni  x\rightarrow x\, \rho (\b) ^{1/d}= x \,r_c(1)/\ell (\b)\in \RR^d .$$
 Since the above function    maps $\hat \Pp _{\rho(\b)}$ onto
$\hat \Pp_1$ and points at distance $r(\b)$ into points at distance
$r_c(1)/2$, the random variable
\begin{equation}\label{couperin}
\hat \omega  _\b \left( W_{r(\b)} \left( B_{r(\b)}\right )[ \hat \omega _\b]  \right)
\end{equation}
has the same law as
\begin{equation}\label{bolero}
 \hat \omega_*  \left( W_{r_c(1)/2} \left( B_{ r_c(1)/ 2}\right) [ \hat\omega_*]  \right)\,,
 \end{equation}
where $\hat \omega _* $ has law $\hat \Pp _1$. The random variable
(\ref{bolero}) is $\b$--independent and has finite moments due to
 Lemma \ref{boundedmom}. Hence all moments of (\ref{couperin}) are $\b$--independent and finite. Due to (\ref{musica}), (\ref{viola}) is satisfied and we can apply Proposition \ref{ricetta}.

\qed


\section{Proof of Theorem \ref{teo-Mottup} in the general case}\label{general_case}

We now explain how one can derive Theorem \ref{teo-Mottup} in the
general (non Poissonian) case, under the domination assumption (ii).

First we observe that  due to assumption (ii)  $\hat \Pp$ has finite
moments $\rho_\k$ for all $\k \geq 0$. In fact, assumption
(ii)  trivially 
implies that  $\hat \Pp^{(p)}$ has finite moments. Hence,
denoting by $X_n$ a generic binomial variable with parameters $n,p$, we have 
\begin{multline*}
\infty>   \EE_{\hat \Pp ^{(p)} } \left ( \hat \xi ([0,1]^d )^\k
\right)= \sum _{n=1}^\infty  \hat \Pp \bigl( \hat \xi ([0,1]^d )=
n\bigr)\sum _{j=0}^n
j^\k \binom{n}{j} p^j (1-p)^{n-j}=\\
  \sum _{n=1}^\infty \hat \Pp
\bigl( \hat \xi ([0,1]^d )= n\bigr) \EE ( X_n^\k )\,.
\end{multline*}
 Since for any positive  integer $\k$ we have
 $\EE ( X_n^\k )= c(\k) n^\k $, this
implies that
$$ 
\rho_\k = \sum _{n=1}^\infty \hat \Pp \bigl( \hat \xi ([0,1]^d )=
n\bigr) n^\k <\infty\,, \qquad \forall \k \in \NN\,.
$$


Define
\begin{equation}\label{davide1}
E(\b): = \b ^{ -\frac{d}{\a +1+d } }\, . \end{equation} Since 
$E(\b)\downarrow 0 $ as $\b\uparrow \infty$, due to (\ref{cond_nu})
we can find $\b_*$ such that
$$ \g:= \nu \bigl( [-E(\b_*),E(\b_*)]\bigr) \leq p\,.
$$
Since the $\g$--thinning $\hat \Pp ^{(\g)}$ is stochastically
dominated by the $p$--thinning $\hat \Pp ^{(p)}$, assumption (ii)
 in Theorem \ref{teo-Mottup}  remains valid with $p$ replaced
by $\g$. Hence, without loss of generality, we can assume that
$\g=p$ in assumption (ii), i.e.
\begin{equation}
p=\nu \bigl( [-E(\b_*),E(\b_*)]\bigr) \end{equation} for some
$\b_*$.

  In what follows we take $\b \geq  \b_*$ and define
\begin{align}
& \ell (\b):=   \lambda \b ^{\frac{\a+1}{\a+1+d}} \,, \label{davide2}\\
&   \Ee ^\b (\xi):= \left\{
 \{x,y\}\,:\, x,y\in \hat \xi\,,\, x \not = y\,,\, |E_x|\leq E(\b),\, |E_y|\leq E(\b)\,   \text{ and } |x-y|\leq \ell (\b) \right\}\,
 , \label{davide3}
\end{align}
where the positive   $\b$--independent  constant $\lambda$ will be
fixed at the end.

We want to apply Proposition \ref{ricetta} where,  given $\Ee^\b
(\xi)$, the graph $\Gg ^\b (\xi) =\bigl (\Vv ^\b (\xi), \Ee ^\b
(\xi) \bigr)$ is defined by  (\ref{chitarra}).
  Trivially, the set $\Ee
^\b (\xi)$  satisfies (\ref{tamburo}) and  (\ref{covariante}), and
condition (A1) is   fulfilled. Moreover,
 due to (\ref{uuu}),
$$
\label{debussybis} \{x,y\}\not\in \Ee^\b (\xi)   \Rightarrow
c_{x,y}(\xi)\leq \exp\left\{-\ell(\b) \wedge [\k_1 \b E(\b)]\right
\}\leq \exp \left\{
 -c \,
 \b ^{   \frac{\a+1 }{\a +1+d }   }   \right \}=:C(\b)\,,
$$
for some $\b$--independent positive constant $c$.   Therefore,
condition (A3) is satisfied. Moreover,   as already observed,
$\rho_\k<\infty$ for all $\k>0$. Hence, in order to obtain the bound
(\ref{montebianco}), which corresponds to (\ref{eq-Mott}), we only
need to verify (\ref{viola}). We will prove that
 \begin{equation}\label{losannina}
\limsup _{\b\uparrow \infty } \EE_{\Pp_0}\left(\vert
C_{0}^\b(\xi)\vert^3\right)<\infty.
\end{equation}
Due to Campbell identity (\ref{campbellbis}), which holds also with
$Q_K$ replaced by $\Lambda_K(0)$ \cite{DVJ}, we can write
$$
 \EE_{\Pp_0}
 \left(
 \vert C_{0}^\b(\xi)\vert^3
 \right) =\frac{1}{\rho K^d}
 \EE_{\Pp} \left(
\int _{\Lambda _K(0)} \hat \xi (dx) \vert C_{0}^\b(S_x
\xi)\vert^3\right)\,,
$$
where $\Pp $ denotes the $\nu$--randomization of $\hat \Pp$.

Let us define the conditioned measure 
$$ \nu_*:= \nu \bigl( \cdot \,|\, |E_0|\leq E(\b_*) \bigr)\,.
$$
Then
$$
 \nu \bigl( [-E(\b), E(\b)]\bigr) =  \, p \,\nu_*\bigl( [-E(\b),
E(\b)]\bigr)\,.
$$
Hence,   for $\b\geq \b_*$, the random set   $ \bigl\{x\in \hat
\xi\,:\, |E_x|\leq E(\b) \bigr\}$ with $\xi$ chosen with law $\Pp$
 and the random set   $ \bigl\{x\in \hat
\xi\,:\, |E_x|\leq E(\b) \bigr\}$ with $\xi$  chosen with law
$\Pp_*$, defined as the $\nu_*$--randomization of $\Pp ^{(p)}$, have
 the same distribution.    Since the graph $\Gg^\b(\xi)=\bigl( \Vv ^\b
 (\xi), \Ee^\b (\xi) \bigr)$  is univocally determined by  the set $ \bigl\{x\in \hat
\xi\,:\, |E_x|\leq E(\b) \bigr\}$, we conclude that
\begin{equation}\label{bruna1}
 \EE_{\Pp_0}
 \left(
 \vert C_{0}^\b(\xi)\vert^3
 \right) =\frac{1}{\rho K^d}
 \EE_{\Pp_*} \left(
\int _{\Lambda _K(0)} \hat \xi (dx) \vert C_{0}^\b(S_x
\xi)\vert^3\right)\,.
\end{equation}
In order to bound the r.h.s. of (\ref{bruna1}) using the domination
assumption (ii), we consider the partition  of $\RR ^d$ in the cubes
$\Lambda _K(x)$, $x\in K \ZZ^d$, and  to each  $A \subset \RR^d$ we
associate  the sets
\begin{align*}
& V_K(A) :=\bigl \{x\in K \ZZ^d\,:\,\Lambda _K(x) \cap A
\not = \emptyset\bigr\}\,,\\
& E_K(A) :=\left\{ \, \{x,y\} \,:\, x,y \in V_K(A) \,,\; x\not = y
\,,\; |x-y|\leq \ell (\b ) + d \sqrt{K} \right\}\,.
\end{align*}
We define $S_K(A)$ as the connected cluster in the graph $G_K(A)=
\bigl( V_K(A), E_K(A) \bigr)$ containing the origin if $0\in V_K(A)
$, and as the empty set if $0\not \in V_K(A)$. Finally, we set
$$
C_K(A):= \cup _{x\in S_K(A)}\left( \Lambda _K(x) \cap A \right)\,.
$$
If    $A=\bigl \{x\in \hat \xi\,:\, |E_x|\leq E(\b)\bigr\}$ we will
simply write $ \Cc^\b_K(\xi)$ for $ C_K(A)$.

\begin{lemma}
For all $x\in \Lambda _K(0)$, it holds
\begin{equation}\label{saturno1}
 C^\b_0 (S_x \xi) + x \subset \bigl(\xi \cap \Lambda _K(0)\bigr)\cup  \Cc ^\b _K (\xi)\,.
\end{equation}
\end{lemma}
\begin{proof}
Due to the covariant property (\ref{covariante}), $ C^\b_0 (S_x \xi)
+ x = C_x ^\b (\xi)$.  If  this set is empty the thesis is trivially
true. Otherwise suppose that  $z\in C_x ^\b (\xi)$. If $z \in
\Lambda_K(0)$ then   $z$ belongs to the r.h.s. of (\ref{saturno1}).
If $z \in \Lambda_K(u)$ for some $u\in K\ZZ^d \setminus \{0\}$, by
following the path connecting $x$ to $z$ in $\Gg ^\b (\xi)$ we can
define a sequence of  distinct points $u_0=0, u_1, u_2, \dots ,
u_{n-1}, u_n=u $ in $K\ZZ^d $ such that for each $i$, $0\leq i \leq
n-1$, there exist  points $a_i \in\Lambda_K (u_i)$, $b_i \in \Lambda
_K (u_{i+1})$ with $\{a_i, b_i\}\in \Ee^\b (\xi)$.  Hence $\{u_i,
u_{i+1} \}\in E_K(A) $ for all $0\leq i\leq n-1$, where $A= \{x\in
\hat \xi\,:\, |E_x|\leq E(\b)\}$. This proves that $0$ and $u$ are
connected in the graph $G_K (A )$. In particular, $u\in S_K (A)$ and
therefore $z\in \Lambda_K(u)\cap A \subset \Cc ^\b _K(\xi)$. \qed
\end{proof}

Due to the above lemma   we get the bound
\begin{multline}\label{bruna2}
  \text{r.h.s. of (\ref{bruna1}) } \leq \frac{c}{\rho K^d}
 \EE_{\Pp_*} \left(
\int _{\Lambda _K(0)} \hat \xi (dx) \left(\hat\xi(\Lambda_K(0) ) ^3
+\vert \Cc^\b _K (\xi)\vert^3\right)\right)=\\
c \,p \,\EE_{\Pp_*}   \left(\hat\xi(\Lambda_K(0)
 ) ^3  \right)+ c\,
 p \,\EE_{\Pp_*} \left( \vert \Cc^\b _K (\xi) \vert^3 \right)\,.
\end{multline}
  Since $$
\EE_{\Pp_*}   \left(\hat\xi(\Lambda_K(0)
 ) ^3  \right)= \EE_{\Pp ^{(p)} } \left(\hat\xi(\Lambda_K(0)
 ) ^3  \right)<\infty\,,
 $$
 in order to conclude the proof we only need to prove that
 \begin{equation}\label{viaggiare}
 \limsup_{\b\uparrow \infty}
 \EE_{\Pp_*} \left( \vert \Cc^\b _K (\xi) \vert^3 \right)<\infty\,.
\end{equation}

Let us derive from the domination assumption (ii) that
\begin{equation}\label{bruna3}
\EE_{\Pp_*} \left( \vert\Cc^\b _K (\xi)\vert^3 \right) \leq
\EE_{\Pp_{*, \rho'} } \left(\vert \Cc^\b _K (\xi) \vert^3 \right)\,,
\end{equation}
where $\Pp _{*, \rho'}$ is the $\nu_*$--randomization of the Poisson
point process $\hat \Pp _{\rho'}$. To this aim, we define
$$\Phi   _K (\hat \xi) =\left\{\hat  \xi \bigl( \Lambda _K(x)\bigr) \,:\, x\in K \ZZ^d
 \right\} \,, \qquad \hat\xi \in \Nn \,.
$$
We claim  that, given a  marked point process
 $\Qq$ obtained as $\nu_*$--randomization of a stationary
simple point process, the conditional expectation
\begin{equation}\label{fuga}
   \EE_{\Qq}\left(\vert \Cc^\b _K (\xi) \vert^3 \;\big|\; \Phi_K \right)
\end{equation}
 is an increasing function in $\Phi _K$ that does not depend on
$\Qq$.   
In order to prove this statement, we write  $\text{Bin}(N,p)$ for a
generic binomial variable with parameters $N,p$ and recall that
 $\text{Bin}(N,p)$ is stochastically dominated by  $\text{Bin}(N',p)$
if $N\leq N'$. Given  $\Phi_K(\hat\xi)$, the random variables
 $\bigl(a_x (\xi),\, x\in K\ZZ^d \bigr)$ defined as
$$
a_x (\xi) = \bigl| \{z\in \hat \xi\cap \Lambda_K(x) \,:\, |E_z|\leq
E(\b)\} \bigr|
$$
are independent binomial r.v.'s with parameters  $\hat \xi(\Lambda_K (x)  )$,
$ \nu_*[-E(\b), E(\b)]$. In particular,  the conditional law of 
 $\bigl(a_x (\xi),\, x\in K\ZZ^d \bigr)$ given $\Phi_K$ does not depend on 
$\Qq $ and it  increases with
$\Phi_K$. Besides, the cardinality  $\bigl|\Cc^\b _K (\xi)\bigr|$
         is an increasing function of $\bigl(a_x (\xi),\, x\in K\ZZ^d \bigr)$.
Since the composition of increasing functions is increasing, we conclude that 
the conditional expectation (\ref{fuga}) is an increasing function of 
$\Phi_K$, independent of $\Qq$. This proves our claim.



\smallskip

 Since by assumption (ii) the random field $\Phi_K(\hat \xi)$
with $\hat \xi$ chosen with law $\hat \Pp ^{(p)}$ is stochastically
dominated by the random field $\Phi_K(\hat \xi)$ with $\hat \xi$
chosen with law $\hat \Pp _{\rho'}$, we obtain that
\begin{multline*}
  \EE_{\Pp_*} \left( \vert\Cc^\b _K (\xi)\vert^3 \right) =
\EE_{\hat \Pp ^{(p)} } \left(\, \EE _{\Pp_*} \left(  \vert\Cc^\b _K
(\xi)\vert^3|   \Phi _K (\hat \xi )\right)\,\right)\leq \\
\EE_{ \hat \Pp_{\rho'} } \left(\, \EE _{ \Pp_{*, \rho'}  } \left(
\vert\Cc^\b _K (\xi)\vert^3|   \Phi _K (\hat \xi )\right)\,\right)=
\EE_{\Pp_{*,\rho'}} \left( \vert\Cc^\b _K (\xi)\vert^3 \right)\,,
\end{multline*}
thus concluding the proof of (\ref{bruna3}).

\smallskip

  Due to (\ref{bruna3}), in order to derive (\ref{viaggiare})
and complete the proof of Theorem \ref{teo-Mottup} we only need to
show that
\begin{equation}\label{bruna4}
  \limsup_{\b\uparrow \infty }\EE_{\Pp_{*, \rho'} }
\left(\vert \Cc^\b _K (\xi) \vert^3 \right)<\infty\,.
\end{equation}
To this aim we will use scaling and percolation arguments as in the
previous section.

\smallskip

 The random set    $A= \left\{ x\in \hat \xi \,:\, |E_x|\leq E(\b) \right\}$, where $\xi$ is chosen with
law $\Pp _{*,\rho'}$, is a Poisson point process with intensity
$$
\mu (\b) = \rho' \nu_*\bigl( [-E(\b), E(\b)] \bigr)\,.
$$
 Hence, by definition of $\Cc ^\b _K(\xi)$, we have 
\begin{equation}\label{pasqua1}
  \EE_{\Pp_{*, \rho'} } \left(\vert
\Cc^\b _K (\xi) \vert^3 \right) =\EE_{\hat \Pp _{\mu(\b)} }
\left(\vert C_K (\hat \xi ) \vert^3 \right)\,.
\end{equation}
Let
$$ r(\b):=2 \sqrt{d} \ell (\b)/ 3\,.$$
Recall that $B_s$ denotes the closed ball of radius $s$ centered at
the origin.  Using the same notation as in  the previous section, for
$\b$ large enough  we can bound
\begin{equation}\label{pasqua2}
  \EE_{\hat \Pp _{\mu(\b)} }  \left(\vert C_K (\hat \xi )
\vert^3 \right)\leq \EE_{\hat \Pp _{\mu(\b)} }
\left[ \hat\xi\left( W_{r(\b)} (B_{r(\b)}) [\hat \xi] \right)^3
\right]\,.
\end{equation}
By the  scaling invariance of the Poisson point process, for each
$\g>0$,
$$
\EE_{\hat \Pp _{\mu(\b)} }
\left[ \hat\xi\left( W_{r(\b)} (B_{r(\b)}) [\hat \xi] \right)^3
\right]=
\EE_{\hat \Pp _{\mu(\b)}\g ^d  }
\left[ \hat\xi\left( W_{r(\b)/\g } (B_{r(\b)/\g }) [\hat \xi]
\right)^3 \right]\,.
$$
Taking
$$
\g : = \mu (\b) ^{-1/d}= \left(\rho ' \nu_* \bigl( [-E(\b),
E(\b)]\right)^{-1/d}   \, ,
$$
we get
\begin{equation}\label{cesi}
\EE_{\hat \Pp _{\mu(\b)} }
\left[ \hat\xi\left( W_{r(\b)} (B_{r(\b)}) [\hat \xi] \right)^3
\right]=
\EE_{\hat \Pp _1}
\left[ \hat\xi\left( W_{r(\b)/\g } (B_{r(\b)/\g }) [\hat \xi]
\right)^3 \right]\,.
\end{equation}
Due to the definition of $\nu_*$, $E(\b)$  and assumption
(\ref{cond_nu}) we get that
$$
\g \geq c\, \b^{\frac{\a+1}{\a+1+d}} $$ for some positive constant
$c$ depending only on $p,\rho', d$ and on the constant $c_0$
appearing in (\ref{cond_nu}). Since $\ell (\b)= \lambda \b^{
\frac{\a+1}{\a+1+d}}$, then
\begin{equation}\label{romani}
 r(\b)/\g= 2\sqrt{d} \ell (\b) / (3 \g) \leq  2 \sqrt{d} \ell (\b)\b^{-
\frac{\a+1}{\a+1+d}}/(3c)  = 2\sqrt{d} \lambda /(3c) \,.
\end{equation}
It is enough to choose $\lambda $ such that
$$  2 \sqrt{d}\lambda /(3c) \leq  r_c(1)/2\,.
$$
With this choice  (\ref{cesi}) and (\ref{romani})  imply that
\begin{equation}\label{pasqua3}
 \sup_{\b>0} \EE_{\hat \Pp _{\mu(\b)} }
\left[ \hat\xi\left( W_{r(\b)} (B_{r(\b)}) [\hat \xi] \right)^3
\right]\leq \EE_{\hat \Pp _1}
\left[ \hat\xi\left( W_{r_c(1)/2 } (B_{r_c(1)/2  }) [\hat \xi]
\right)^3 \right]\,.
\end{equation}
The  r.h.s. in the above inequality is finite due to Lemma
\ref{boundedmom}.  This observation together with 
(\ref{pasqua1}), (\ref{pasqua2}) and (\ref{pasqua3})  imply
(\ref{bruna4}). This concludes  the proof of Theorem
\ref{teo-Mottup}.

\begin{rem}
In the case that $\hat\Pp$ is stochastically dominated by the Poisson 
point process $\hat \Pp_{\rho '}$ one can give a much simpler 
proof of Theorem  \ref{teo-Mottup}, that we describe in what 
follows.  Denote by $\Pp$ and   $ \Pp_{\rho '}$ the $\nu$--randomization  of
 $\hat \Pp$ and  $\hat \Pp_{\rho '}$, respectively. Due to the definition of
stochastic domination given in Section \ref{basicfacts}, one can exhibit a coupling between   $\Pp$ and   $ \Pp_{\rho '}$ such that $\xi\subset\xi'$ almost surely,  with $(\xi, \xi')$
 denoting the random sets with marginal
distributions given by  $\hat \Pp$ and  $\hat \Pp_{\rho '}$, respectively.

 Consider the graph
$\Gg ^\b (\xi)$ introduced in  Section  \ref{poisson_case}. Due to Campbell
identity  (\ref{campbellbis}), we can write
$$
  \EE_{\Pp_0}\left(\vert
C_{0}^\b(\xi)\vert^3\right)= \frac{1}{\rho}  
\EE_{ \Pp} \left ( F_\b(\xi)\right) \,, \qquad F_\b(\xi):=\int _{Q_1}\hat\xi (dx)
 \left(\vert
C_{0}^\b(S_x \xi)\vert^3\right)
.
$$
 Since $F_\b(\xi)\leq F_\b(\xi ')$ if $\xi \subset\xi '$ and 
 due to the above coupling   between   $\Pp$ and   $ \Pp_{\rho '}$, 
 we can conclude that 
$$ 
\EE_{ \Pp} \left ( F_\b(\xi)\right)\leq 
\EE_{ \Pp_{\rho'} } \left ( F_\b(\xi)\right)\,.$$
Due to Campbell identity  (\ref{campbellbis}),   the r.h.s. in the 
above expression equals $ \rho ' \EE_{\Pp_{0,\rho'}}\left(\vert
C_{0}^\b(\xi)\vert^3\right)$ which, 
as proven  in Section \ref{poisson_case},    is bounded from above uniformly in $\b$. Hence  we have that
\begin{equation}\label{mr}
\sup_{\b>0}   \EE_{\Pp_0}\left(\vert
C_{0}^\b(\xi)\vert^3\right)\leq \frac{\rho'}{\rho}\sup_{\b>0} 
 \EE_{\Pp_{0,\rho'}}\left(\vert
C_{0}^\b(\xi)\vert^3\right )  <\infty\,.
\end{equation}
At this point it is enough to apply Proposition  \ref{ricetta}: condition (\ref{viola}) is fulfilled due to (\ref{mr}),
 while all other conditions can be easily checked.
\end{rem}

\smallskip

\section{Point processes with uniform bounds on the local density}\label{WS}

Let us   prove that the conditions of Theorem \ref{teo-Mottup}
are fulfilled by stationary point processes with uniform bounds:

\begin{prop}\label{salmone}
Let $\hat\Pp$ be a stationary simple point process such that, for suitable
positive constants $K$ and $N$,
\begin{equation}
\hat \xi \bigl( \Lambda _K(x)\bigr) \leq N \,,\qquad \forall x \in 
K \ZZ ^d\,, \qquad
 \hat \Pp \text{ a.s.}
 \end{equation}
where  $ \Lambda _K(x)= x + [-K/2, K/2)^d $.

Then, for each $q\in (0,1]$, the $q$--thinning $\hat \Pp ^{(q)}$ of
$\hat \Pp $ satisfies condition (ii) in Theorem  \ref{teo-Mottup}.
\end{prop}

In order to prove the above statement, we use a standard result of 
stochastic domination (cf. Lemma 1.1 in    \cite{LSS}):
\begin{lemma}
Let $p\in[0,1]$ and 
let $\sigma=\bigl( \sigma _x\,:\, x\in \ZZ^d \bigr)$ be a random field
s.t.  $\sigma_x\in \{0,1\}$ for all $x \in  \ZZ^d$. Suppose that
\begin{equation}
P\bigl( \sigma_x=1\,|\, \sigma =\zeta \text{ on } \ZZ^d \setminus\{x\} \bigr)\leq p\,,
\end{equation}
for almost all $\zeta\in \{0,1\}^{\ZZ^d} $ and for all $x \in\ZZ^d$. Then $\sigma $ is stochastically 
dominated by the  Bernoulli site percolation with parameter $p$, i.e.\
one can define random fields $(\sigma', \omega)$ with 
$$
\sigma ' _x\leq \omega _x \,  \qquad \forall x\in \ZZ^d \,, \qquad\text{a.s. },
$$
$\sigma'$ having the same law of $\sigma$ and $\omega$ being given by 
$\omega=\bigl( \omega  _x\,:\, x\in \ZZ^d \bigr)$ where $\omega _x$ are i.i.d. random variables taking value $1$ with probability $p$ and value $0$ with probability $1-p$.
\end{lemma}

\noindent
{\sl Proof of Proposition \ref{salmone}}.
If $q<1$ set $p=q$, otherwise fix some $p\in (0,1)$. Consider the random 
fields $\sigma  =\bigl( \sigma _x\,:\, x\in \ZZ^d \bigr) $, $\tau
=\bigl( \tau  _x\,:\, x\in \ZZ^d \bigr) $ having value in $\{0,1\}^{\ZZ^d}$ defined as 
\begin{align*}
&  \sigma_x (\hat \xi) =
 \begin{cases} 1 & \text{if } 
 \hat \xi \bigl( \Lambda _K(x) \bigr ) \geq 1 \,, \\
0 & \text{otherwise}\,;
\end{cases}
&  \tau_x (\hat \xi) =
 \begin{cases} 1 & \text{if } 
 \hat \xi \bigl( \Lambda _K(x) \bigr ) \geq N  \,, \\
0 & \text{otherwise}\,.
\end{cases}
\end{align*}
Then, given $\z\in \{0,1\}^{\ZZ^d}$, we have 
\begin{align*}
& \hat\Pp ^{(p)} \bigl( \sigma_x=1\,|\, \sigma =\zeta \text{ on } \ZZ^d \setminus\{x\} \bigr)
\leq 1- (1-p)^N=: p' \,, \\
&  \hat\Pp _{\rho '}  \bigl( \tau _x=1\,|\, \tau  =\zeta \text{ on } \ZZ^d \setminus\{x\} \bigr)= P(Z\geq N)=: \tilde p \,,
\end{align*}
where $Z$ is a Poisson variable with mean  $\rho' K^d$.
Since $p'<1$ we can choose $\rho '$ large enough so 
that $p'\leq \tilde p$. 
Due to the previous lemma, we can conclude that 
 the random field $\sigma$ with $\hat \xi$ chosen with law $ \hat\Pp ^{(p)}$
is stochastically dominated  by 
the site Bernoulli percolation with parameter $\tilde p $, which is stochastically dominated by  the random field $\tau $ with $\hat \xi$ chosen with 
 Poisson law $\hat\Pp _{\rho '}$.
Due to the transitivity of stochastic domination and since 
$$Y(x) \leq N \sigma _x , \qquad N \tau _x \leq  Y (x)\,,\qquad \forall x \in \ZZ ^d \,,
$$
we can conclude that the random field $Y$ with $\hat\xi$ chosen with law 
$ \hat\Pp ^{(p)}  $ is stochastically dominated by the 
 random field $Y$ with $\hat\xi$ chosen with law $\hat\Pp _{\rho '}$.
\qed

\bigskip

\smallskip

As corollary of Proposition \ref{salmone}, we obtain that Theorem \ref{teo-Mottup}
can be applied to crystals or diluted crystals. In order to be more precise, let us start 
with a crystal, i.e. (cf. \cite{AM})  a 
locally finite  set $\Gamma\subset \RR^d$ such that for a suitable basis
$v_1, v_2, \dots, v_d$ of $\RR^d$, it holds
\begin{equation}\label{paranza}
\G -x =\G \, \qquad \forall x\in G:=\bigl \{ z_1v_1+z_2v_2 +\cdots +z_d v_d 
\,:\, z_i \in \ZZ \;\; \forall i  \bigr\}\,.
\end{equation}
Let $\D$ be the elementary cell
$$
\D:=\bigl \{t_1 v_1+t_2 v_2 +\cdots+t_d v_d \,:\, 0\leq t_i <1\;\; \forall  i \bigr\}\,.
$$
Note that both the group  $G$ and the cell $\D$ depend on the basis $v_1, v_2, \dots, v_d$.

Let $\omega=\bigl(\omega_x\,:\, x\in \G\bigr)$ be a site Bernoulli percolation on
$\G$ with parameter $p\in (0,1]$ and let $V$ be a random vector independent of
 $\omega$, chosen
in the elementary cell $\D$ with uniform distribution. Then consider the simple point process
$$
\hat \zeta := \sum _{x\in \G}\omega _x  \d_{V+x}\,,
$$
obtained from the set $\G$ by a spatial randomization and a $p$--thinning, and call $\hat \Pp$
 its law. The following holds:

\begin{prop}\label{mare1000}
The simple point process  $\hat \zeta $  with law $\hat \Pp$ 
is stationary and does not depend on the specific basis
 $v_1, v_2, \dots, v_d$ satisfying (\ref{paranza}). Moreover, its Palm distribution $\hat \Pp _0$ is given by
\begin{equation}\label{sogliola}
\hat\Pp _0 =\frac{1}{|\D\cap \G|} \sum _{u \in \D\cap \G} P _u 
\end{equation}
where $P  _u$ is the law of the simple  point process 
$$
\hat \zeta _u :=\d_0 +  \sum _{x\in \G\setminus \{u\} } \omega _x\d_{x-u}   \,.
$$
\end{prop}
Since the Palm distribution  $\hat \Pp _0$ depends only on $\hat \Pp$, the above 
result implies that  $\hat \Pp _0$ does not depend on the  specific basis
 $v_1, v_2, \dots, v_d$. If $p=1$, the Palm distribution corresponds  to   choosing a point $u$  of the crystal
inside  the elementary cell $\D$ with uniform probability and translating the crystal of $-u$.
 If $p<1$, in addition
to the previous step one erases points different from the origin independently with probability
$1-p$. The resulting simple point process is what we called $p$--diluted crystal obtained from 
 $\G$. Trivially, it has uniform bounds in the local density. Hence, due to Proposition \ref{salmone},
it fulfills the assumptions of Theorem \ref{teo-Mottup}.

\begin{proof}
Due to the   translation invariance (\ref{paranza}) one easily proves that 
  $\hat\Pp$ is stationary. Let us  prove that it does not depend on the specific 
 basis  $v_1, v_2, \dots, v_d$. Given a Borel subset $A\subset \RR^d$
with finite positive Lebesgue measure, we  define 
$\hat\Pp_ A$ as the law of the simple point process $\sum_{x\in\G}\omega_x \d_{W+x}$,  where
$\omega$ is a site Bernoulli percolation on $\G$ with parameter $p$ and $W$ is a vector independent from
$\omega $ and  chosen in $A$ with 
uniform probability. Note that $\hat\Pp =\hat\Pp_\Delta$. 
By means of   (\ref{paranza}) one can easily check that
$\hat \Pp _\D = \hat \Pp _A$ if $A$ is a union of sets of the form $\D+x$, $x\in G$. Now, consider
the elementary cell $\D'$ associated to another basis  $v'_1, v'_2, \dots, v'_d$ satisfying (\ref{paranza}).
By the previous observation,
 $\hat \Pp _{\D'} =\hat\Pp_{N\D'}$ for each  positive integer $N$. 
Define $A_N $ as the union of all the sets of the form $\D+x$, $x\in G$, included in $N\D'$. Then 
$\hat \Pp _\D = \hat \Pp _{A_N}$. Finally observe that, given a bounded measurable function $f:\hat\Nn \rightarrow \RR$, 
it holds
$$
\left|
\EE_{\hat\Pp_{N\D'} } (f) - \EE_{\hat  \Pp _{A_N}
} (f)\right|\leq C(\D,\D', f) /N\,.
$$
Hence the same estimate holds with  $\hat\Pp_{N\D'}$ and
 $\hat  \Pp _{A_N}$ replaced by $ \hat\Pp _{\D'}$ and 
 $ \hat\Pp _{\D}$, respectively. Taking $N\uparrow \infty$, we conclude that  $ \hat\Pp _{\D'}= \hat\Pp _{\D}$. Hence
the law $\hat\Pp$ does not depend on the specific 
 basis  $v_1, v_2, \dots, v_d$ satisfying (\ref{paranza}).

We   prove the characterization (\ref{sogliola}) of the Palm distribution $\hat \Pp _0$ by means
of Campbell identity (\ref{campbell}). Take a bounded    measurable function $f:\hat \Nn _0 \rightarrow \RR$
and denote by  $\rho$ the intensity of $\hat \Pp$. It is simple to check that
\begin{equation}\label{finefine}
\rho =p |\D \cap \G |/|\D|\,,
\end{equation}
where 
$|\D \cap \G |$ denotes the cardinality of $\D\cap \G$ and $|\D|$ denotes the Lebesgue volume of $\D$.
 Since   (\ref{campbell}) holds also with $Q_K$ replaced by $\D$, we get that  
\begin{equation}\label{marina111}
\EE_{\hat \Pp _0} (f) =\frac{1}{\rho |\D| } \EE _{\hat\Pp}\left( \int _{\D } \hat \xi (x) f(S_x\hat  \xi)\right)\,. 
\end{equation}
Let us define $\G_\omega$ as the support of the measure $\sum _{x\in \G} \omega _x \d_x $. Denoting by
$\EE_\omega$ the expectation w.r.t. the site Bernoulli percolation $\omega$ we can write
\begin{equation}  \label{marina222}
\begin{split}
  \EE _{\hat\Pp}\left( \int _{\D} \hat \xi (x) f(S_x \hat \xi)\right)& =
\EE_\omega\left[ \frac{1}{|\D|} \int _\D dy \left( \sum _{z\in \D\cap (\G_\omega +y ) }
f\bigl( \G _\omega  +y -z \bigr) \right)\right] \\
& =
\EE_\omega\left[ \frac{1}{|\D|} \int _\D dy \left( \sum _{u\in (\D-y) \cap \G_\omega  }
f\bigl( \G _\omega  -u \bigr) \right)\right]\\
& =
\EE_\omega\left[ \frac{1}{|\D|} \int _\D dy \left(
 \sum _{u\in (\D-y) \cap \G  } \omega_u
f\bigl(  ( \G _\omega  -u) \cup \{0\}    \bigr) \right)\right]\\
&= p \EE_\omega\left[ \frac{1}{|\D|} \int _\D dy \left( \sum _{u\in (\D-y) \cap \G   }
f\bigl( ( \G _\omega  -u) \cup \{0\}  \bigr)   \right)\right]\\
&= 
   \frac{p}{|\D|}  \int _\D dy  \sum _{u\in (\D-y) \cap \G   }  \EE_\omega
\left[ f\bigl( ( \G _\omega  -u) \cup \{0\}  \bigr) \right]\,.
\end{split}
\end{equation}
Due to (\ref{paranza}), it is simple to check that the last summation does not depend on $y$. Hence
 from the 
above identities (\ref{finefine}),  (\ref{marina111}) and (\ref{marina222}) we get that
\begin{multline}
\EE_{\hat \Pp _0} (f)= \frac{p}{\rho|\D| } 
 \sum _{u\in \D  \cap \G   }  \EE_\omega
\left[ f\bigl( ( \G _\omega  -u) \cup \{0\}  \bigr) \right]=  \frac{p}{\rho |\D|} 
\sum _{u\in \D  \cap \G   } \EE_{ P_u } (f)=\\
\frac{1}{|\D \cap \G|}  \sum _{u\in \D  \cap \G   }
 \EE_{ P_u } (f)\,,
\end{multline} 
thus concluding the proof of (\ref{sogliola}).

\end{proof}

\bigskip

\bigskip

\noindent {\sl Acknowledgment}. One of the author, A.F.,   thanks
the
  Centre de Math\'ematiques et d'Informatique (CMI), Universit\'e    de Provence, for the kind hospitality and   acknowledges the financial support of GREFI--MEFI.


\end{document}